\documentclass[12pt,preprint]{aastex}

\lefthead{Escala et al.}
\righthead{Gravitational Fragmentation in Galaxy Mergers: A Stability Criterion}

\def\msun{M_{\odot}}

\begin{document}

\title{Gravitational Fragmentation in Galaxy Mergers: A Stability
  Criterion.}

\author{Andr\'es Escala, Fernando Becerra, Luciano del
  Valle and Esteban
Castillo}
\affil{Departamento de
  Astronom\'{\i}a, Universidad de Chile, Casilla 36-D, Santiago,
  Chile.}

\begin{abstract}

We study the gravitational stability of gaseous streams in the complex
environment of a galaxy merger, because mergers are known to be places
of ongoing massive cluster formation and bursts of star formation. We
find an analytic stability parameter for case of gaseous streams
orbiting around the merger remnant. We test our stability criteria
using hydrodynamic simulations of galaxy mergers, obtaining
satisfactory results. We find that our criterion successfully predicts
the streams that will be gravitationally unstable to fragment into
clumps.  
\end{abstract}

\section{Introduction}

Galaxy mergers are believed
to be not just common events in the universe, but also fundamental
pieces in the evolution of galaxies since they trigger bursts of star
formation (Larson \& Tinsley 1978; Sanders \& Mirabel 1996) and they
are a key ingredient in the formation Elliptical Galaxies and Bulges
(Toomre \& Toomre 1972; Mihos \& Hernquist 1994, 1996; Kazantzidis et
al. 2005; Di Matteo et al. 2007). More recently, it was also found
that many interacting and merging galaxies are places of current
massive cluster formation (Schweizer 1998; Mengel et al. 2008). 

Standard numerical simulations of galaxy mergers that include gas
(Mihos \& Hernquist 1994; Barnes \& Hernquist 1996; Kazantzidis et
al. 2005; Cox et al. 2006; Di Matteo et al. 2007), have been able to
reproduce the observed starbursts occurred during the merging process
on nuclear gaseous disks. However, they intentionally avoid
fragmentation through high minimum temperatures and large
gravitational softening lengths, therefore, they failed to reproduce
formation of massive star clusters. Only recently, simulations have
the resolution required to study the gas fragmentation on at least
large scales (Bournaud et al. 2008; Saitoh et al. 2009; Teyssier et
al. 2010; Matsui et al 2012). In those simulations, massive star
clusters are indeed formed by gas fragmentation into collapsing clumps
and therefore is relevant to have a criteria for gravitational
instabilities in such complex environment.

The study of gravitational stability of fluids started with the work
of Jeans (1902) for a uniform, infinite and isothermal gas. Later
extended by Bonnor (1956) and Ebert (1955) for a finite and
spherically symmetric fluid, a rotationally supported one (Toomre
1964; Goldreich \& Lynden Bell 1965), a magnetized fluid
(Chandrasekhar \& Fermi 1953), among others. In this letter, we study
the stability of the gaseous streams in the complex environment of a
galaxy merger, by means of Smoothed Particle Hydrodynamics (SPH)
numerical simulations.

This work is organized as follows. We start with a discussion of the
physical processes relevant in stabilizing gaseous streams in galaxy
mergers, with analytical estimates for a stability criterion in \S
2. Section 3 continues discussing the setup of the galaxy mergers
simulations and the resolution needed to resolve
gravitational instabilities in galaxy mergers. In \S 4, we test the
stability criteria by performing hydrodynamical simulations of galaxy
mergers with the resolution discussed in \S3. Finally in \S 5, we
summarize the results of this work.

\section{Basic Physical Ingredients: Gas Pressure and Motion}

High-resolution simulations of galaxy mergers generally found galactic
streams, such as tails and bridges at large scales and more complex
ones on the inner kpc scales, in which collapsing clumps are
ubiquitous features formed by gravitational instabilities. However, no
gravitational instability criterion for the complex and irregular case
of gaseous streams in a galaxy merger has been found.

The most basic physical processes that could overcome gravity in
absence of magnetic and other fields are gas pressure and
motion. Since gas pressure is isotropic, does not depend on the
geometry of the fluid, only on its local density and temperature,
therefore its stabilizing role on small scales is the same that in a
fluid with a regular geometry (i.e with a given symmetry). On the
order hand, the motion of streams is much more complex and not
constrained to a single plane, but on its central regions (inner kpc,
where the bulk of the star and cluster formation happens) is
characterized by streams orbiting around the center of mass of the
newly formed system.

A simple and useful approach is to model individual streams as a piece
of a rotating annulus. In such a case, from vector calculus is known
that the rotational component of its motion is well described by an
angular frequency, which is defined relative to an origin O of the
coordinate system in which we describe the motion:
\begin{equation}
\rm \vec{\Omega}_{o} = \frac{\vec{r} \times  \vec{v}}{\vec{r} \,
  \cdot
  \, \vec{r}} = \frac{\vec{r} \times  \vec{v}}{r^2}  =
\hat{r} \times  \frac{\vec{v}}{r}\,\,\,\, ,
\label{Omega}
\end{equation}
where $\rm \vec{r}=r\,\hat{r}  \,\, and \,\,  \vec{v}$ are,
respectively, the position and velocity vectors.

Under this approximation, the stability of individual streams is a
very similar problem to the stability of annuli in a rotating sheet,
with the difference that the streams in the merger case do not belong
to the same plane ($\rm \vec{\Omega}_{o}$ of individual streams can
have a different magnitude and direction). In such a case, from
dimensional analysis it is straightforward to conclude that results
from standard gravitational instability analysis in a rotating sheet
(Toomre 1964; Goldreich \& Lynden Bell 1965; Binney \& Tremaine 2008)
should still be valid for a given stream: there is a range of unstable
length scales limited on small scales by thermal pressure (at the
Jeans length $\rm \lambda_{\rm Jeans} = C_{\rm s}^2 / G\Sigma_{\rm
  gas} $) and on large scales by rotation (at the critical length set
by rotation, which for this case can be defined as $\rm \lambda_{\rm
  rot} \equiv \pi^2 G \Sigma_{\rm gas} /|\vec{\Omega}_{o}|^2$). All
  intermediate length scales are  unstable, the most rapidly
  growing mode has a wavelength $2 \, \lambda_{\rm Jeans}$ and the
  most unstable mode has a wavelength $\lambda_{\rm rot}/2$. Only a combination of pressure and rotation can stabilize the stream,
this happens when the range of unstable wavelengths shrinks to zero
(i.e. the two scales are comparable) and this occurs for $\lambda_{\rm
  Jeans} \geq (q/\pi)^2 \, \lambda_{\rm rot}$ (Escala \& Larson
2008)
. Therefore a stream will be stable if:
\begin{equation}
\rm |\vec{Q}_{o}| \equiv  \frac{ C_{S}  \,  |\vec{\Omega}_{o}| }{G \,
  \Sigma_{gas}}  \ge q \,\,\,\, ,
\label{Q}
\end{equation}
where $\rm C_{S}$ is the gas sound speed,  $\rm  |\vec{\Omega}_{o}|$
is the  norm of the angular frequency vector, G is the gravitational
constant,  $\rm \Sigma_{gas}$ is the gas surface density and q is a
number of the order of unity. Otherwise, if $\rm |\vec{Q}_{o}| < q$, a
stream will be unstable.  

It is important to point out that the concept of angular
frequency depends on the origin of the coordinate system chosen to
describe the motion, and contrary to the case   of the rotating sheet,
there is not an obvious single choice for all streams in a galaxy
merger. This becomes relevant in section 4, when we compare
the results of this section with numerical simulations, for testing if the 
stability criteria given by Eq $\ref{Q}$ is
valid or not. Our approach in section 4 will be to check if with a single
origin of the coordinate system O, Eq $\ref{Q}$ is able to predict the
gravitational instability of the streams. This assumption will introduce
 changes in the value of the angular frequency and therefore, in the
 determination of the value for the threshold q (will be an average
 value for all streams). However, our aim is to have a simple criteria that
 can be easily applied be other authors and in such a case, is better
 to have a single $\rm |\vec{Q}_{o}|$ with an average fitting parameter
 q, than one $\rm |\vec{Q}_{o\alpha}|$ and $\rm q_\alpha$ for each
 $\rm \alpha$th stream.  

In the case that all the streams are coplanar and orbit around the
same point, we recover the standard Toomre Q parameter for a rotating
sheet (=$\rm C_{S}  \, \Omega/ G \, \Sigma_{gas}$; Toomre 1964),
since now the direction of $\rm \vec{\Omega}_{o}$ and the origin O
chosen to describe the motion, is the same for all streams. 

\section{Simulations of gravitational fragmentation in galaxy mergers}

In the following we perform a set of idealized numerical experiments
aimed to test if the stability parameter  (Eq. \ref{Q}), successfully
predicts the streams in a galaxy merger  that will be gravitationally
unstable to fragment into clumps. These experiments are constructed as
simple as possible, in order to guarantee that the gas fragmentation
is only due to gravitational instabilities. For that reason, we use an
isothermal equation of state instead of having a multiphase medium and
do not include any feedback processes from star formation and/or AGN
which will make more complex the analysis as it includes new sources that
may trigger fragmentation. Without including this extra physics we
cannot aim to have a realistic description of the ISM, but will be
enough for our main purpose, which is to study the onset of
gravitational instability at large scales in a galaxy merger.   


The simulation consists on the merger of two equal mass disk galaxies
and we let the two galaxies collide in a parabolic orbit with
pericentric distance $\rm R_{min}$ =7.35 kpc. The simulations start
with an initial separation of 49 kpc, where the separation distance is
measured between the mass centers of the two galaxies and the initial
inclination angle between disk planes of individual galaxies is $\rm
90^{o}$. The galaxies are initialized using the code GalactICS, in
particular we used their  `Milky Way model A' (see Kuijken \& Dubinski
1995 for details). In each galaxy model, we include a gaseous disk
with the same exponential profile as the stellar component (Kuijken \&
Dubinski 1995) and with a total gas mass corresponding to the 10\% of
the total stellar disk mass. The gas has an isothermal equation of
state, $\rm P = c_{S}^{2} \, \rho$, where the sound speed is fixed at
$\rm c_{S}= 12.8 \, km s^{-1}$, corresponding to a gas temperature of
$\rm \sim 2 \times 10^{4}$K. At t=0, the gaseous disk of each isolated galaxy is gravitationally stable. In our simulations, we use the following
internal units : [Mass] = $5.8 \times 10^{11}\msun$, 
[Distance] = 1.2 kpc and G=1. The total number of particles is
420,000, being 200,000 for sampling the gas, 120,000 for the dark
matter halo, and 80,000 for the disk component and 20,000 for the
bulge.  

The simulations were evolved using the SPH code Gadget-2 (Springel
2005), up to a time t=160  (in internal time units), which correspond
to a point where the galaxies are after their third (and final)
pericentric passage and in which most of the gas ($>$ 80\%) has been
fueled to the central kpc. Fig 1 (a, b, c, d) show the evolution of
the system at four times t = 32 (a), 54 (b), 120 (c) and 136 (d) which
corresponds to before (a) and after (b) the first pericentric passage,
second pericentric passage (c) and in their third pericentric
encounter (d). The ring/oval structures seen on  Fig. 1(a, b, c), are ubiquitous features since early simulations of galaxy mergers (e.g., Schwarz 1984) and are believed to be tightly wound spirals that are the gas response to tidal forcing (e.g., Barnes \& Hernquist 1996). 
 

\subsection{Minimum Gravitational Resolution}

Before analyzing the stability of gaseous streams it is necessary to
check that we have the gravitational resolution required to resolve
the fragmentation of streams into collapsing clumps, for that reason
we performed a convergence test. We restarted the original simulation
with a gravitational softening length $\rm \epsilon_{soft}$=0.4 at
t=132, with the following gravitational softening $\rm
\epsilon_{soft}$: 0.04, 0.02, 0.01, 0.008, 0.006 in internal units. 

Fig 1 (e, f, g, h) shows the evolution of the restarted simulation at
a later time t=134, in a region of radius 2 internal distance units
(2.4 kpc) for different gravitational softening lengths: 0.4(e),
0.04(f), 0.01(g), 0.006(h).  Fig 1 (e, f, g, h) shows that as the
softening lengths decrease, we find more gas fragmentation until a
point in which the resulting simulations converge. We find convergence
of the results for $\rm  \epsilon_{soft} \le 0.01$ and for that
reason, we choose to use in the following section a gravitational
softening length of  $\rm \epsilon_{soft} = 0.01$.

The convergence can be understood if we take into account that over
99.6\% of the particles in such region fulfill the condition  $\rm
\lambda_{rot}  \ge 4 \, \epsilon_{soft}$  for $\rm  \epsilon_{soft} =
0.01$ and below. The convergency when $\rm \lambda_{rot}$ is resolved
for all particles, is the first suggestion for supporting our
definition for $\rm  \lambda_{rot} $ in this environment with
disordered motion ($\rm \lambda_{\rm rot} \equiv \pi^2 G \Sigma_{\rm
  gas} /|\vec{\Omega}_{o}|^2$). 
This convergence is an evidence that
the minimum requirement  to resolve  fragmentation at least on the
largest scales, is to be able to resolve gravity below  our
definition for the largest unstable scale $\rm \lambda_{rot}$.

In this set of numerical experiments, we resolve fragmentation from
the largest unstable scale down to our gravitational resolution. Below
the gravitational softening, sub-fragmentation is artificially damped
but our aim is to study if Eq 2 can predict the instability of streams
and for that purpose, is not required to resolve all the range of unstable
wavelengths and is enough with the largest one, since $\rm
\lambda_{rot}$ is the first unstable wavelength to appear (i.e. the most unstable mode; Binney \& Tremaine 2008).


This resolution test illustrates that in a set of simulations with the same temperature, fragmentation
 can be prevented just by the gravitational resolution. This contradics the interpretation of Teyssier et al. (2010), in which the onset of fragmentation is always  asociated with a decrease in the temperature. However, Teyssier et al. (2010) changes both temperature and resolution, being unable to disentangle which one (or both) is responsible for the onset of fragmentation. This reinforces our approach of testing the stability criteria (Eq. 2) against a set of simple simulations, where the variation of parameters can be fully controlled.

Finally, it is worth to mention that in this section and in the rest
of the paper, we will focus only in the
 fragmentation of the gaseous component. The reason is that the stellar component behaves approximately as an adiabatic
fluid (i.e. the kinetic energy in stellar motions cannot be lost or "radiated" away from the merging system). This yields to a rapid conversion of coherent motions into random ones during the merger, with the subsequent increase of 
the velocity dispersion in the stellar component, stabilizing the
stellar system against runaway fragmentation.


\section{Test of the stability criteria $\rm |\vec{Q}_{o}|$}

After checking the gravitational resolution needed to resolve fragmentation at least
on scales below those of the largest collapsing clumps, we will focus
on testing the stability criteria discussed in \S 2
(Eq. \ref{Q}). Since most of the streams in the inner 2.4 kpc of the
system already fragments for T $\rm \sim 2 \times 10^{4}$K, we will
perform a set of simulations in which we increase the temperature and
see how some streams become stable. We will check if the criteria
given by Eq $\ref{Q}$, successfully predicts if a stream should be
stable or not.

Fig 2 (a, b, c) shows the evolution of the gas density for the
system restarted  with a  gravitational softening length of  $\rm
\epsilon_{soft} = 0.01$, at t=132, for different temperatures T=$\rm 2
\times 10^{4}(a),  2 \times 10^{5}(b) \, and
\,10^{6} \,K(c)$ and evolved to a later time t=133.2.  The comparison
between different temperatures (a to c in Fig 2) clearly shows that
more streams become stable as we increase the temperature.

Fig 2 (d, e, f) shows  $\rm |\vec{Q}_{o}|/q$ computed for each
particle at the time in which all the simulations were restarted with
$\rm \epsilon_{soft} = 0.01$ (t= 132), for different temperature
T=$\rm 2 \times 10^{4}(d),  2 \times 10^{5}(e) \,
and \,10^{6} \,K(f)$. For computing $\rm \vec{\Omega}_{o}$, 
we choose  as origin O of the coordinate system the total center of mass (G) of the merging
galaxies, because the system as a whole orbits around G and is also an
inertial point for an isolated merger (in absence of external
forces). From Eq. \ref{Q}, the threshold for stability should be
around   $\rm |\vec{Q}_{o}|/q = 1$, which corresponds to the yellow
particles in the figure 2, the green and blue particles should be
unstable and the red ones stable. 

The direct comparison of two sides of Fig 2 (a-d, b-e and c-f
pairs), shows overall a good agreement between the predicted unstable
streams (showed in green and blue in Fig 2 (d, e, f)) and the
streams that eventually fragments on the corresponding Fig 2 (a, b, c). In particular, the bluest stream in Fig 2 d (and in light green in
Fig 2 f) is the most unstable region and the only one that strongly
fragments in all simulations (including Fig 2 c) besides the increase
in temperature up to $\rm 10^{6} \,K$. 

We find that the stability of gaseous streams is better described for
a threshold value q$\sim$0.4, in fact, we plot $\rm |\vec{Q}_{o}|/0.4$
in Fig 2 (d, e, f). This is approximately a factor of 2 lower than
the value expected for a uniformly rotating isothermal disk (q = 1.06;
Goldreich \& Lynden-Bell 1965). However, it is important to emphasize
that the actual value of q should depend on the origin O chosen for
the coordinates system.  

In order to check the numerical reliability of our results, we
performed the same fragmentation runs with 2,000,000  particles for the
different temperatures showed in Fig 2 (a, b, c). Fig 2 (g, h, i) shows the evolution of the gas density for the high-resolution
runs. By direct comparison of two sides of Fig 2 (a-g, b-h, and c-i pairs), we found differences in the low-density regions, as
expected in the SPH technique, and slightly different positions in some
streams, also expected due to the different granularity of the
gravitational potential. However, we found the same results in terms of the onset of fragmentation on streams (i.e. which
ones fragments at a given temperature and which ones don't) and in the
number of clumps formed. Therefore, for the purpose of our study, we
found consistent results between the low and high resolution
runs. This supports the reliability of our numerical experiments that tests Eq. 2. 

Finally, it is important to note that although we successfully tested Eq. 2 for
t=132, it should be valid at any given time for the value of the
angular velocity vector in such moment (only with small variations on
the threshold q). This is particularly important  because, during the
evolution of a galaxy merger, the properties of  any stream  (angular
frequency vector, surface density, etc.) can drastically  change on a timescale
comparable to a crossing time. 
To check that our criteria is valid at any time, we restarted 
the original low-resolution simulation again at a different
time  t=137.2, with a gravitational softening $\rm
\epsilon_{soft} = 0.01$ and a temperature T=$\rm 2 \times 10^{4} K$.

Fig 3a shows the evolution of the gas density for the
system restarted at t=137.2 and evolved to a later time  t=138.4. Fig
3b shows  $\rm |\vec{Q}_{o}|/q$ computed for each particle at the time
t=137.2 using the same  previously used threshold value q$=$0.4. The direct
comparison of two sides of Fig 3, shows again an overall  good
agreement between the predicted unstable streams (showed in green and
blue in Fig 3b) and the streams that eventually fragments in the
evolution of the SPH run (Fig 3a).

Although  the gas properties (and predicted $\rm |\vec{Q}_{o}|$) in
this second restarting time  has considerably changed compared to the
properties in the original ones (Fig 2),  Fig 3 shows that the  $\rm
|\vec{Q}_{o}|$ computed at the restarting time t=137.2, successfully
predicts  which  streams  will fragment and which  ones don't.
 This is besides some minor discrepancies that may arise if we focus on 
some small scale features showed in Fig 3. For example, a careful 
inspection of Fig 3a shows a smooth spiral pattern around  a big clump (slightly up 
from the center of the image). On the corresponding Fig 3b, both features
are in blue ($\rm |\vec{Q}_{o}|/q <<1$) which is expected for the big clump but not 
for the smooth spiral feature.

Fig 4 shows a zoom-in of such region, in which the left panel shows
the gas density and the right panel the corresponding $\rm
|\vec{Q}_{o}|/q$. From Fig 4 is straightforward to realize that the spiral is composed by  only few tens of particles that were lost within  the several thousands of particles
that compose a clump (i.e. less than 1\%). These particles, were probably lost  thru the interactions with other clumps and in fact the spiral pattern ends on the closest collapsed clump, suggesting that are particles lost  during strong gravitational interactions between the clumps. These minor discrepancies are inherent of the complexity of the problem, because processes that happens on the subsequent evolution, such as clump-clump interactions, are of course not included on this or any stability criterion.

In order to quantify these minor discrepancies, we plot in Fig. 5 the surface density of each particle against  $\rm
|\vec{Q}_{o}|/q$. The left panel of Fig. 5, shows the surface density
at the restarting time t=132 against $\rm |\vec{Q}_{o}|/q$ computed at
the same time. The  dashed line represents $\rm \Sigma_{gas} \propto
(|\vec{Q}_{o}|/q)^{-1}$, which is the overall trend of particles at the
restarting time. This is expected from the definition given
by Eq 2, taking into account that $\rm C_{S}$ is constant and that the
dispersion from the overall trend, is due to  variations of $\rm |\vec{\Omega}_{o}|$ among particles.

The middle and right panels of Fig. 5, shows the surface density
at the  t=133 (middle) and 134 (right) against $\rm |\vec{Q}_{o}|/q$ computed at
the  restarting time t=132. By construction, the particles on middle and
right panels of Fig. 5 can only move in the vertical direction (relative to their position in the left panel) and therefore it will inherently introduce scatter into the  $\rm \Sigma_{gas} \propto
(|\vec{Q}_{o}|/q)^{-1}$ trend, since the surface density and $\rm |\vec{Q}_{o}|/q$ are computed at two different times. Beside the increase in scatter, we see
a coherent vertical change for $\rm |\vec{Q}_{o}|/q \leq 1$ as the
collapse proceeds. For isothermal simulations like the ones carried out
in this work, once the gravitational instability is started it will
proceed until the particles reach separations comparable to the
softening length and this behaviour happens regardless the  $\rm
|\vec{Q}_{o}|/q$ or $\rm \Sigma_{gas}$ values. The horizontal
sarturated region on middle and right panels of Fig. 5 (at $\rm
log(\Sigma_{gas}) \geq 0$) denotes
this behaviour. On the other hand, the time to move to the saturated
region (free fall time) it does depends on  $\rm \Sigma_{gas}$ and
this explains why dark region of particles with $\rm log(\Sigma_{gas}) \leq 0$
recedes towards higher  $\rm |\vec{Q}_{o}|/q$  as times evolves
(from $\rm log(|\vec{Q}_{o}|/q) \geq -1.5$ in the middle panel to $\rm log(|\vec{Q}_{o}|/q) \geq -1$ in right panel of Fig. 5).

In the right  panel of Fig. 5, the particles in the region $\rm
log(\Sigma_{gas}) \leq 0$ and $\rm log(|\vec{Q}_{o}|/q) \leq -1$, are
representative of particles lost from the collapsing clumps, like the
ones showed in the spiral feature of Fig. 4. In the same way, the
particles in the region $\rm log(\Sigma_{gas}) \geq 0$ and $\rm
log(|\vec{Q}_{o}|/q) \geq 0$,  represents particles from stable regions that are
gravitationally captured by the collapsing clumps. Besides these
departures, the overall trend is clear and is showed in the right
panel of Fig. 5, where a drastic change at  $\rm
log(|\vec{Q}_{o}|/q) = 0$ is clearly seen and that is due to  the collapse of the
unstable regions, which  
moves their particles vertically up, producing the vertical saturated region seen for $\rm log(|\vec{Q}_{o}|/q) \leq 0$.

\section{Summary}

In this paper, we have studied the gravitational stability of gaseous
streams in the complex environment of a galaxy merger, using
hydrodynamic simulations. 

We find that the standard Toomre Q stability parameter can be
generalized for case of gaseous streams orbiting around the merger
remnant, by using  the angular frequency vector of each stream. This
is valid as long as the orbital motion of a stream can be well
approximated by the rotational motion around the center of
gravity on a given plane, which is what happens in the inner regions
of the merger remnant.

We test our generalized stability criteria, $\rm |\vec{Q}_{o}| \geq
q$, using  SPH numerical simulations specially  designed for that
purpose. We find that this criteria successfully predicts the streams
that will be gravitationally unstable to fragment into clumps. We find
that the stability of streams is better described  choosing a
threshold value  q$\sim$0.4.  

The generalization of $\rm \lambda_{rot}$ in a galaxy merger, is also
relevant for the formation of massive globular-type  clusters since
its associated mass  $\rm M_{\rm rot} = \Sigma_{\rm gas} \,
(\lambda_{\rm rot}/2)^2$,  is related to  the characteristic mass of
the most massive clusters that are able to form (Escala \& Larson 2008;
Shapiro et al. 2010) and has a role in the triggering of star
formation, since it correlates with the galactic star formation rate
(Escala 2011).  

The numerical validation of stability for $\rm |\vec{Q}_{o}| \geq
q$  opens new possibilities for
future research. One is to apply the  criterion given by Eq. 2 
 to observations of gas-rich galaxy mergers and also, to
simulations with a more realistic description for 
the ISM, that includes feedback processes from star formation and/or AGN.
Another interesting possibility is  to
study when in the evolution of a merger, you have
a larger portion of the gaseous mass with $\rm |\vec{Q}_{o}| \leq
q$ and then be able to determine when the streams would fragment more 
vigorously. 

A.E. acknowledges partial support from the Center for
Astrophysics and Associated Technologies CATA (PFB 06), FONDECYT Iniciacion
Grant 11090216. F.B. and L. del V.  acknowledge  support from Programa
Nacional de Becas de Posgrado (Grant D-22100632 and Grant
D-21090518). The simulations were performed using the HPC clusters
Markarian (FONDECYT 11090216), Geryon (PFB 06) and Levque.



\begin{figure}[h!]
\begin{center}
\includegraphics[width=12cm]{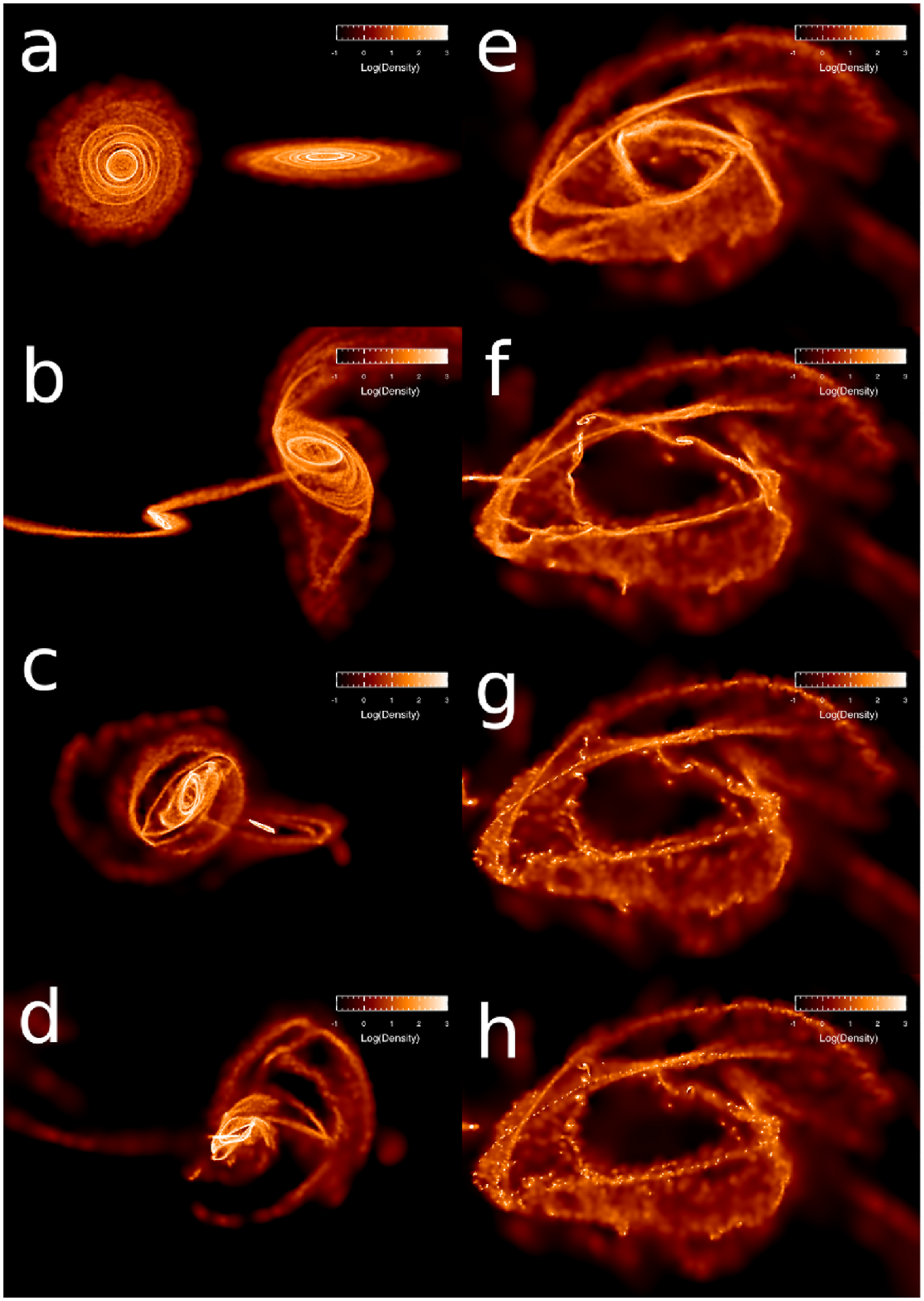}
\caption{Gas density distribution during the evolution of the galaxy merger
  showed in a logarithmic scale. Left side  panels in the figure
  show the  density  distribution in a box of side 25, in internal
  units, at the following times t = 32 (a), 54 (b), 120 (c) and 136
  (d). Right side panels show a zoon in of the
  simulation restarted at t=132 (box of side 4 in internal distance units), evolved up to 
time t=134  using the following gravitational softening lengths:
0.4(e), 0.04(f), 0.01(g), 0.006(h).}
\label{c1p3}
\end{center}
\end{figure}
\vspace{-0.5cm}

\begin{figure}[h!]
\begin{center}
\includegraphics[width=17cm]{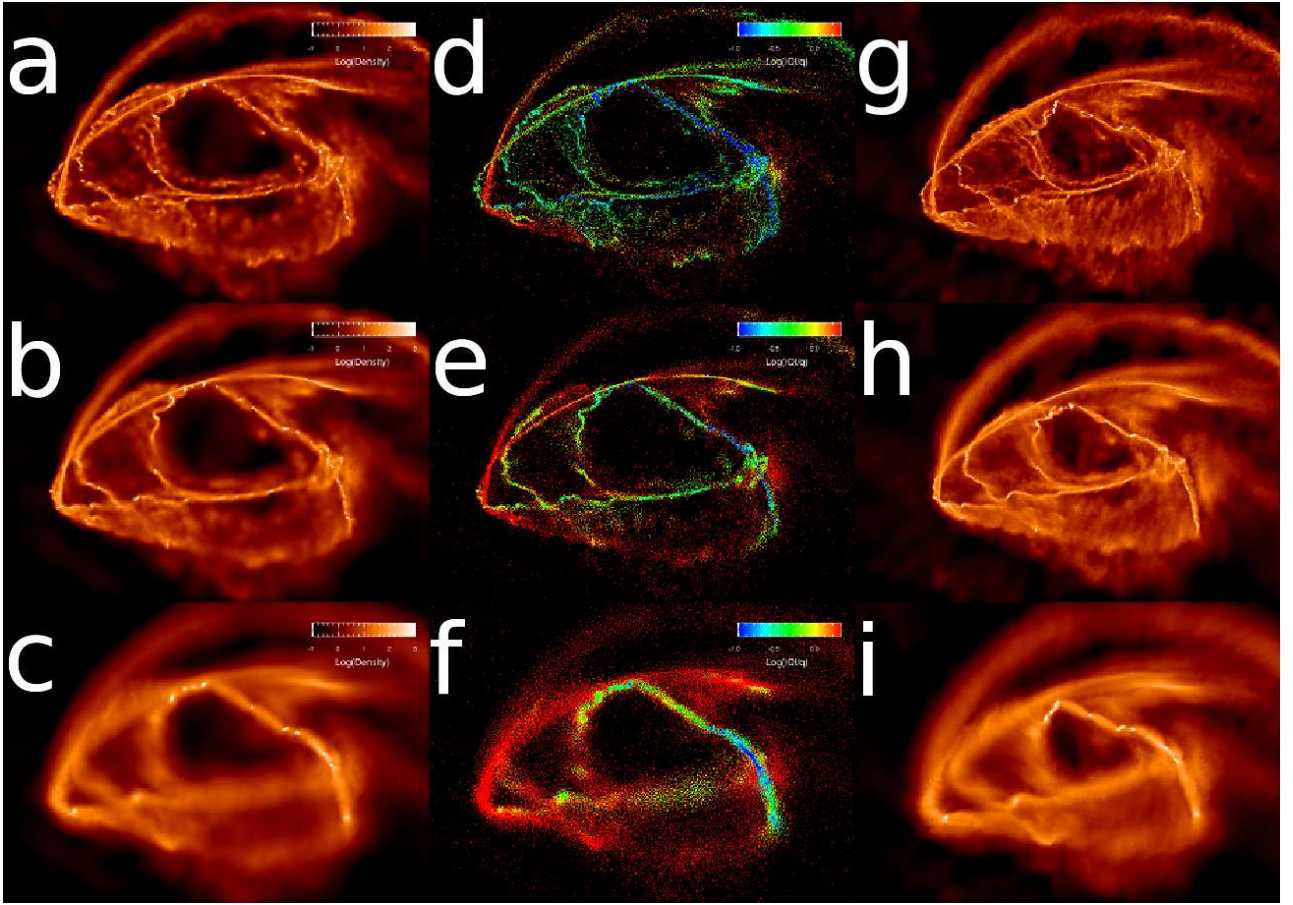}
\caption{Left side  panels   show the evolution of the gas density
  distribution at t=133.2, for different 
temperatures T=$\rm 2 \times 10^{4}(a),  2 \times 10^{5}(b)   \, and \,10^{6}
\,K(c)$.  Middle panels show the predicted  $\rm |\vec{Q}_{o}|/q$ for
each particle, which is computed at the restarting time t= 132, for the following
temperatures T=$\rm 2 \times 10^{4}(d),  2 \times 10^{5}(e) \,and \,10^{6} \,K(f)$. Right side  panels show the evolution of the gas density
  in the high resolution runs, for different temperatures T=$\rm 2 \times 10^{4}(g),  2 \times 10^{5}(h)   \, and \,10^{6} \,K(i)$. In all figures, the boxes have a side of 4 
internal distance units.}
\label{c1p3}
\end{center}
\end{figure}
\vspace{-0.5cm}

\begin{figure}[h!]
\begin{center}
\includegraphics[width=17cm]{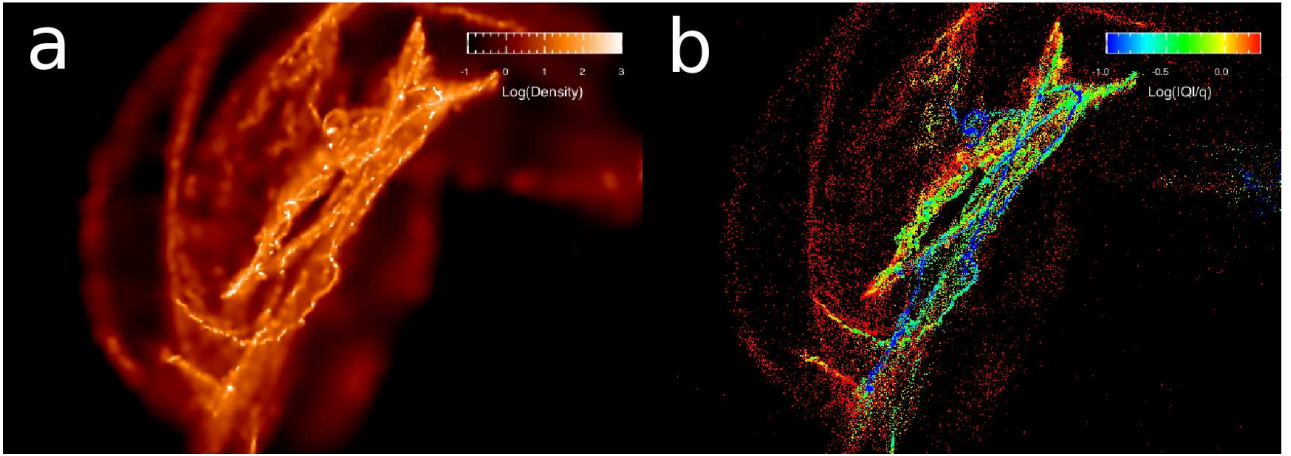}
\caption{a) The gas density distribution at t=138.4 of a simulation
  with a temperature T=$\rm 2 \times 10^{4}$ K, that was restarted at
  t= 137.2  with a gravitational softening $\rm
\epsilon_{soft} = 0.01$. b) The predicted  $\rm |\vec{Q}_{o}|/q$ for
each particle, which is computed at the restarting time t= 137.2, for a
temperature T=$\rm 2 \times 10^{4}$ K.}
\label{c1p3}
\end{center}
\end{figure}
\vspace{-0.5cm}

\begin{figure}[h!]
\begin{center}
\includegraphics[width=17cm]{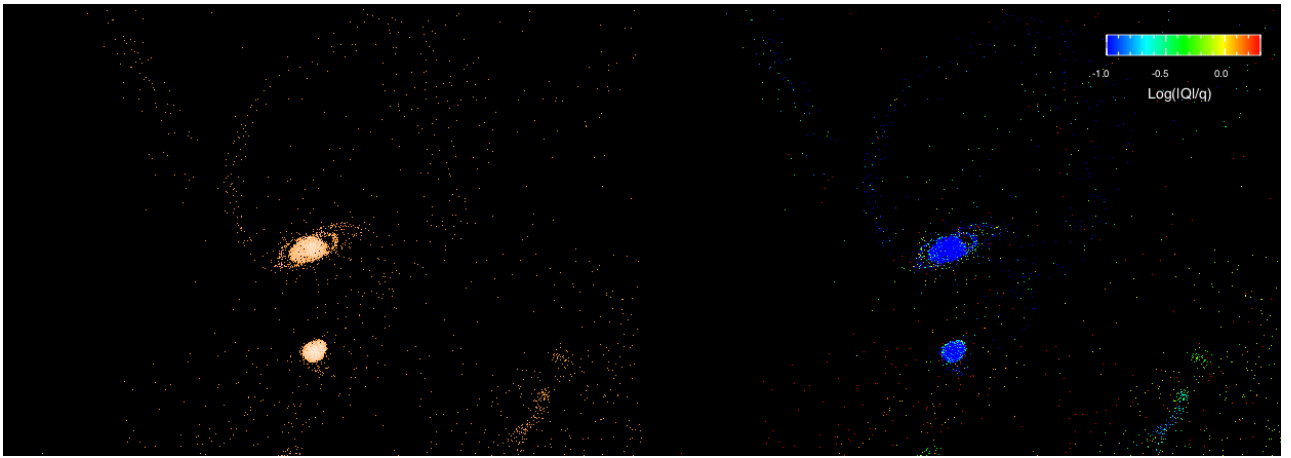}
\caption{Zoom into the simulation showed in figure 3. Left: The gas density for
each particle at t=138.4. Right: The predicted  $\rm |\vec{Q}_{o}|/q$ for
each particle, which is computed at the restarting time t= 137.2}
\label{c1p3}
\end{center}
\end{figure}
\vspace{-0.5cm}

\begin{figure}[h!]
\begin{center}
\includegraphics[width=17cm]{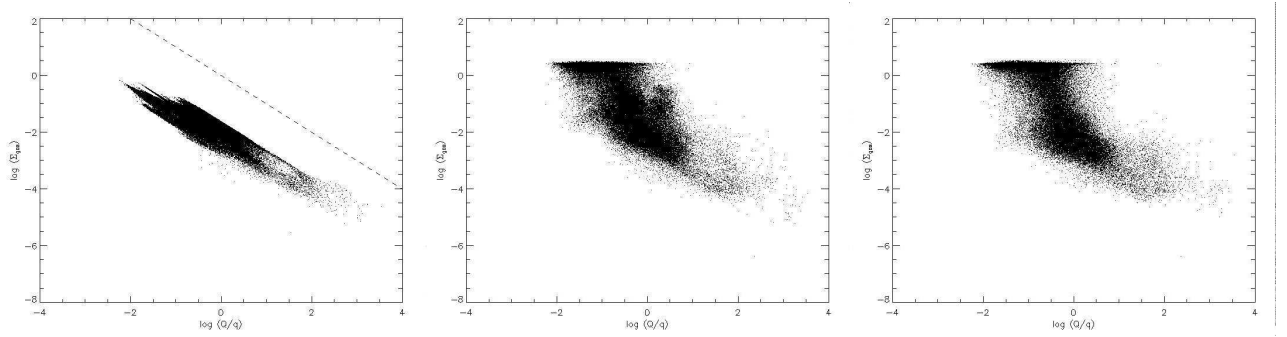}
\caption{Evolution of the surface density as fragmentation proceeds. The predicted  $\rm |\vec{Q}_{o}|/q$ for
each SPH particle, which is computed at the restarting time t= 132, plotted against the surface density of each particle at a time t= 132 (right), 133 (middle) and 134 (left).
}
\label{c1p3}
\end{center}
\end{figure}
\vspace{-0.5cm}




\end{document}